\documentclass{article}
\usepackage{graphics}
\usepackage{psfig}
\usepackage{emulateapj}
\usepackage{apjfonts}
\newcommand{\email}[1]{\authoremail{#1}}
\usepackage[dvips]{color}

\newcommand{\realfigure}[2]{ 
             \hbox{~}\centerline{\psfig{file=#1,angle=0,width=3.5in}}  
             \vspace{0pt}\noindent{\small \addtolength{\baselineskip} 
             {-3pt} \hspace*{0.3cm} {#2} 
             \addtolength{\baselineskip}{13pt}} 
             \vspace{0.1in}}

\usepackage{apjfonts}

\input{psfig.tex}  
\newcommand{\gsim}{\;\lower.6ex\hbox{$\sim$}\kern-7.75pt\raise.65ex\hbox{$>$}\;}
\newcommand{\lsim}{\;\lower.6ex\hbox{$\sim$}\kern-7.75pt\raise.65ex\hbox{$<$}\;}

\begin{document}

\title{RR Lyrae variables in the globular clusters of M31: a first detection of
likely candidates\altaffilmark{1}}

\author{Gisella Clementini}
\affil{Osservatorio Astronomico di Bologna, 
Via Ranzani 1, 
I-40127 Bologna, Italy; gisella@bo.astro.it}
\email{gisella@bo.astro.it}

\author{Luciana Federici}
\affil{Osservatorio Astronomico di Bologna, 
Via Ranzani 1, 
I-40127 Bologna, Italy; luciana@bo.astro.it}
\email{gisella@bo.astro.it}

\author{Carlo E. Corsi}
\affil{Osservatorio Astronomico di Roma, 
Via di Frascati 33, 
I-00040 Monte Porzio Catone, Italy; corsi@coma.mporzio.astro.it} 
\email{corsi@coma.mporzio.astro.it}

\author{Carla Cacciari}
\affil{Osservatorio Astronomico di Bologna, 
Via Ranzani 1, 
I-40127 Bologna, Italy; cacciari@bo.astro.it}
\email{cacciari@bo.astro.it}

\author{Michele Bellazzini}
\affil{Osservatorio Astronomico di Bologna, 
Via Ranzani 1, 
I-40127 Bologna, Italy; bellazzini@bo.astro.it}
\email{bellazzini@bo.astro.it}

\and

\author{Horace A. Smith}
\affil{Department of Physics and Astronomy, Michigan State 
University, East Lansing, MI 48824-1116; smith@pa.msu.edu}
\email{smith@pa.msu.edu}

\altaffiltext{1}{Based on observations made with the NASA/ESA Hubble Space 
Telescope,
obtained from the data archive at the Space Telescope Science Institute. STScI
is operated by the Association of Universities for Research in Astronomy, Inc.
under NASA contract NAS 5-26555.}


\begin{abstract}

The purpose of this paper is to show that RR Lyrae variables exist and can 
be detected in M31 globular clusters. 
We report on the first tentative identification of RR Lyrae
candidates in four  globular clusters of the 
Andromeda galaxy, i.e. G11, G33, G64 and G322.
Based on HST-WFPC2 archive observations in the F555W and F814W filters 
spanning a total 
interval of about 5 consecutive hours we find evidence for  
2, 4, 11 and 8  RR Lyrae variables of both {\it ab} and {\it c} 
Bailey types in G11, G33, G64 and G322, respectively. 
Several more candidates can be found by relaxing slightly the selection 
criteria. 
These numbers are quite consistent with the horizontal branch morphology 
exhibited by the four clusters, starting from the very blue HB in G11, and 
progressively moving to redder HBs in G64, G33 and G322. 
\end{abstract}

%

\section{introduction}
\label{s_intro}
RR Lyraes are the most common class of intrinsic variable stars 
in the Galaxy and
in the Local Group. Their almost constant average absolute magnitude makes them 
primary distance indicators in the Galaxy and in the Magellanic Clouds, 
hence they are cornerstones of cosmological distance and time scales.

Evidence for the presence of RR Lyrae variables has been 
found in almost all of the Local Group 
galaxies where a search has been pushed deep enough to reach the 
horizontal branch (HB).
M31 is no exception to this general rule. However, 
very few RR Lyrae variables have been identified so far in this galaxy.
Pritchet \& van den Bergh (1987) found 30 probable 
field RR Lyraes with the CFH 3.6-m telescope in conditions of 
``excellent" seeing. Difference imaging of M31 has also revealed some 
possible field RR Lyraes (Sugerman, Uglesich \& Crotts, 1999). 
The faintness of the targets and  the severe crowding 
conditions have prevented any further search for RR Lyraes in that galaxy. 
Nevertheless, the estimated relative frequency of such variables 
indicates that a very rich harvest of them should be present
in the stellar spheroid of M31 (Pritchet \& van den Bergh, 1987). 
As for RR Lyraes in M31 globular clusters, no observation has been 
attempted from the ground because the high spatial resolution 
provided so far only by the HST is mandatory to resolve these stars.

Identification of the RR Lyrae population in the globular clusters of M31, 
however, can provide fundamental insights on 
the distance scale, the stellar content, and the formation process of M31, 
since a globular cluster stellar population has the advantage 
over the field of being homogeneous in age and metallicity.

The pulsational properties of RR Lyrae stars are known 
well enough to help in understanding several important questions: 
(i) What is the Population II distance to M31 and how does it compare to 
the distance derived from Population I indicators, Cepheids in particular? 
(ii) What is the relation between the M31 halo and its globular clusters,  
and how do the cluster characteristics compare with their counterparts
in the Milky Way?
(iii) What are the stellar content and the metallicity of the M31 spheroid?

Item (i) is particularly important as a fundamental step to 
strengthen the calibration of the distance scale, hence the determination 
of the time scale and all related cosmological parameters.   
In fact, thanks to its large distance any projection 
and/or line of sight depth effects are much less important in M31 than 
in nearer Local Group systems, such as the Magellanic Clouds. 
Moreover, an Sb I-II giant spiral galaxy provides a much more 
appropriate local counterpart to the Distance Scale Key Project galaxies 
than does the LMC (see reviews 
by Freedman (2000) and Tammann (2000) and references therein). 
Therefore, in any respect except for ease of 
observations, M31 is a much more appropriate cornerstone for the 
distance scale 
than the LMC.  

Item (ii) directly bears upon the halo formation processes in 
M31. Almost all Galactic globular clusters which contain
significant numbers of RR Lyrae stars can be placed into
 Oosterhoff groups I and II (Oosterhoff 1939) on 
the basis of the mean periods and
relative proportions of their RRab and RRc stars.
That is not true for all globular clusters, e.g. those of the LMC, which can 
have properties intermediate between the two Oosterhoff groups
(Bono, Caputo \& Stellingwerf, 1994).
It has been suggested that Milky Way globulars 
belonging to different Oosterhoff groups may represent the products
of different accretion/formation events in the halo (e.g. van den Bergh 1993). 
The existence or absence of the Oosterhoff phenomenon among
the M31 globular clusters may therefore provide an additional constraint on
the similarity of halo formation processes in the two systems.
Finally, a direct comparison of the properties of variables in globular
clusters and in the adjacent fields would provide a very interesting diagnostic
as to whether the spheroid and the globular cluster system share a 
common origin.

In this letter we report on the first tentative identification of RR Lyrae
variable candidates in four globular clusters observed  with 
the HST, 
namely  G11, G33, G64 and G322 (Sargent et al. 1977). 
A larger number of candidate variable objects of unclassified type has also 
been found.

\section{Mining the HST data archive on the Globular Clusters of M31}
\label{s_mining} 
The globular cluster population of the Andromeda galaxy
has been the subject of a number of HST programs.
HST color magnitude diagrams (CMDs) reaching one magnitude below the HB 
level are presently 
available for 19 of the M31 globulars (Ajhar et al. 1996;  Rich et al. 1996; 
Fusi Pecci et al. 1996; Holland, Fahlman \& Richer 1997; Corsi et al. 2000; 
Rich et al. 2001).
Among these clusters we have selected those with: a) well defined CMDs and 
populous enough HBs; and b) a large enough number (n$\geq$4) of 
suitably long individual exposures (in each passband) so that a 
variability search could be performed on the basis of the scatter of the 
individual photometric measures with respect to the average value.

Four clusters satisfy  the above requirements, 
namely G11, G33, G64 and G322 (see Table~1).
All of them are metal poor or of intermediate metallicity, 
thus by
analogy to the Galactic clusters we may expect they contain RR Lyrae stars.
Indeed, at [Fe/H] $\le$ --1.7, G11, G33 and G64 are the sort of clusters which 
in the Milky Way would likely belong to OoII.
G322 is of intermediate metallicity, similar to clusters which would 
likely be OoI systems in the Milky Way.

The positions of the four clusters with respect to the nucleus of M31 are
shown in Figure 1, and projected distances from the M31 center are given 
in column 3 of Table 1. 


\begin{center}
TABLE~1. Information on the four selected clusters
\begin{tabular}{l c c c c}
& & & & \\
\tableline
\tableline
Id. & V & r & E(B--V) & [Fe/H]\\
 &(int.)&($\arcmin$)&(mag) & (spectr.)\\
\tableline
G11 &  16.36 & 75.75 &0.10 & --1.89 \\
G33 &  15.48 & 57.59 &0.22 & --1.74 \\
G64 &  15.08 & 25.35 &0.17 & --1.81 \\
G322&  15.61 & 61.80 &0.10 & --1.21 \\
\tableline
\end{tabular}
\end{center}
\scriptsize
NOTE. - Adopted values of reddening are from Frogel et al. (1980); 
metallicities are from Huchra et al. (1991), positions with respect to the 
center of M31 and integrated 
luminosities are from Corsi et al. (2000).
\normalsize

Four F555W (V) and four F814W (I) PC frames are available for each of the
 selected clusters. 
For each cluster, individual exposures are 20 minutes long  
and consecutive observations cover a total time interval of 5h 15m,  
with the first 2-hr block devoted to V and the remaining time to I. 
These integration times are adequate for our purposes, as they are long 
enough to give an r.m.s. 
error of $\sim$ 0.10 mag on the {\it individual} measures at the HB level, 
and yet short enough to avoid blurring of the light curves.

\section{Analysis of the photometric data: variable search}
\label{s_varsearch}

The F555W and F814W frames for the 
selected clusters were reduced individually using the ROMAFOT reduction 
package, which is especially optimized for accurate photometry in crowded 
fields (Buonanno et  al. 1983). 

\realfigure{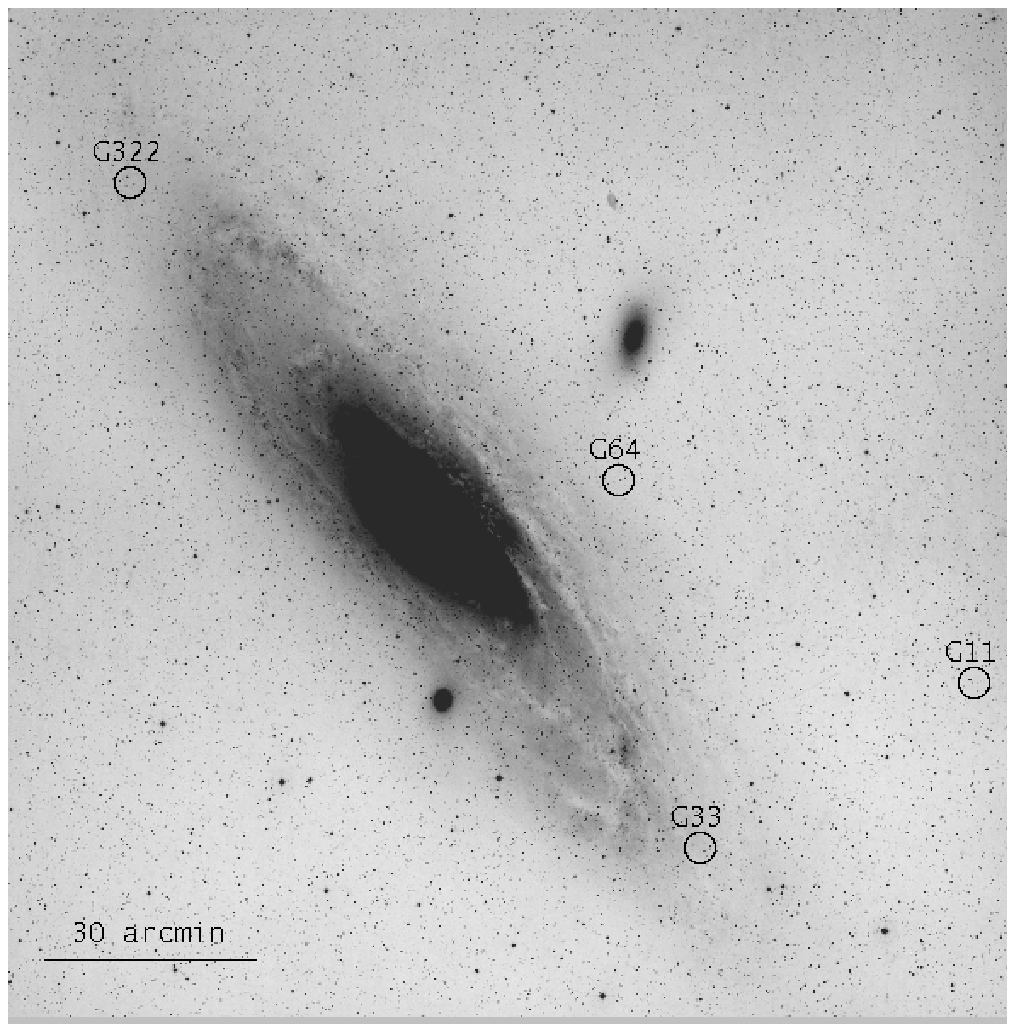}{FIG.~1.\
Location of G11, G33, G64 and G322 with respect to M31 main body.}

We have adopted the following criteria for identifying RR Lyrae-like  
candidates:

i) average luminosity of the available data points within $\pm$ 1 mag of 
the HB average magnitude level at the color of the instability strip. 
No restriction was put on the stellar color, however,  because the 
$< {\rm V}> - <{\rm I}>$ colors of 
variable objects are artificially distorted by phase 
mismatch of the V and I  data that were taken sequentially and not 
simultaneously. 
As a matter of fact, given the time baseline of the data only variables with 
periods shorter than about 2-3 hours have a full coverage in phase in each 
passband, so that the average values of the present data would actually 
correspond to their average magnitudes and colors.   
Vice versa, variables of longer periods may well fall 
outside the boundaries on either side of the instability strip.

ii)  We considered objects with no less than 3 measurements in each  of the 
two photometric passbands, and r.m.s. deviations larger than 0.3 mag., 
i.e. 3$\sigma$ from the average r.m.s. value, in at least one passband. 
This search was performed on the F555W and F814W frames independently.
   
The sequences of individual measurements were then displayed 
as a function of the Julian Day of observation, searching  
for trends, shapes and possible periodicities in the light variations 
which were compatible with RR Lyrae type variations. 

It should be remembered that given the short time coverage of the archival 
data a strong bias exists against the detection of variables with periods 
longer than $\sim$ 5 hours. Thus the only  {\it ab}-type variables which 
can be detected from the present data are 
those caught during the rapid rise to maximum light or the upper part of 
the descending branch in at least one of the two photometric bands. 
On the other hand, detection of the {\it c}-type variables, while favoured 
by the shorter periods, is made difficult by the smaller amplitude
of their light variations.

iii) The final and decisive criterion was a check that the detected 
indication of 
variability was indeed consistent with real RR Lyrae light curves in the V and 
I bands. To this purpose we have considered a grid of V and I light curves 
of RR Lyraes from the globular cluster M3 ([Fe/H]$\sim -1.5$ and 
E(B--V)=0.0, Carretta et al. 1998) sampling 
4 c-type variables with periods in the range 0.29-0.49 d, and 15 ab-type
variables with periods in the range 0.46-0.66 d. These template light curves 
were transferred into the V,I vs. JD plane using their periods, 
were reddened (A$_{\rm V}$=3.1E(B--V) and A$_{\rm I}$=1.9E(B--V) - Schlegel 
et al. 1998) using the appropriate reddening value for each M31 cluster 
(see Table 1), and were converted into HST magnitudes using the relations of 
Holtzman et al. (1995). A comparison could then be performed directly in the 
HST photometric system with the V,I data of the selected candidate 
variable stars: rigid shifts (i.e. the same value for the V and I data) were 
applied vertically 
(because of the different apparent magnitude of stars and templates) and 
horizontally to each star in search of the best match of the individual 
V and I data sets with the most  
suitable template. These shifts of course vary from star to star, because 
they depend on the (unknown) phases and magnitudes at which the stars were 
observed. The stars we eventually selected as likely RR Lyrae 
candidates are those for which at least 3 data points per color fall on the 
respective template light curve within 0.1 mag, i.e. 1$\sigma$. 
Examples of these ``matches'' are shown in Figure 2.
Of course the present data are not sufficient for an independent period 
determination.


These criteria are rather severe, and we can miss candidates because of
occasional larger errors on individual data points, in addition to the 
observational bias that causes us to miss the variables in those phase 
intervals where the light variation is modest. 
According to these criteria, we were 
able to derive a meaningful indication of variability for  2, 4, 11 
and 8 such candidates in G11, G33, G64 and G322, respectively. 

\realfigure{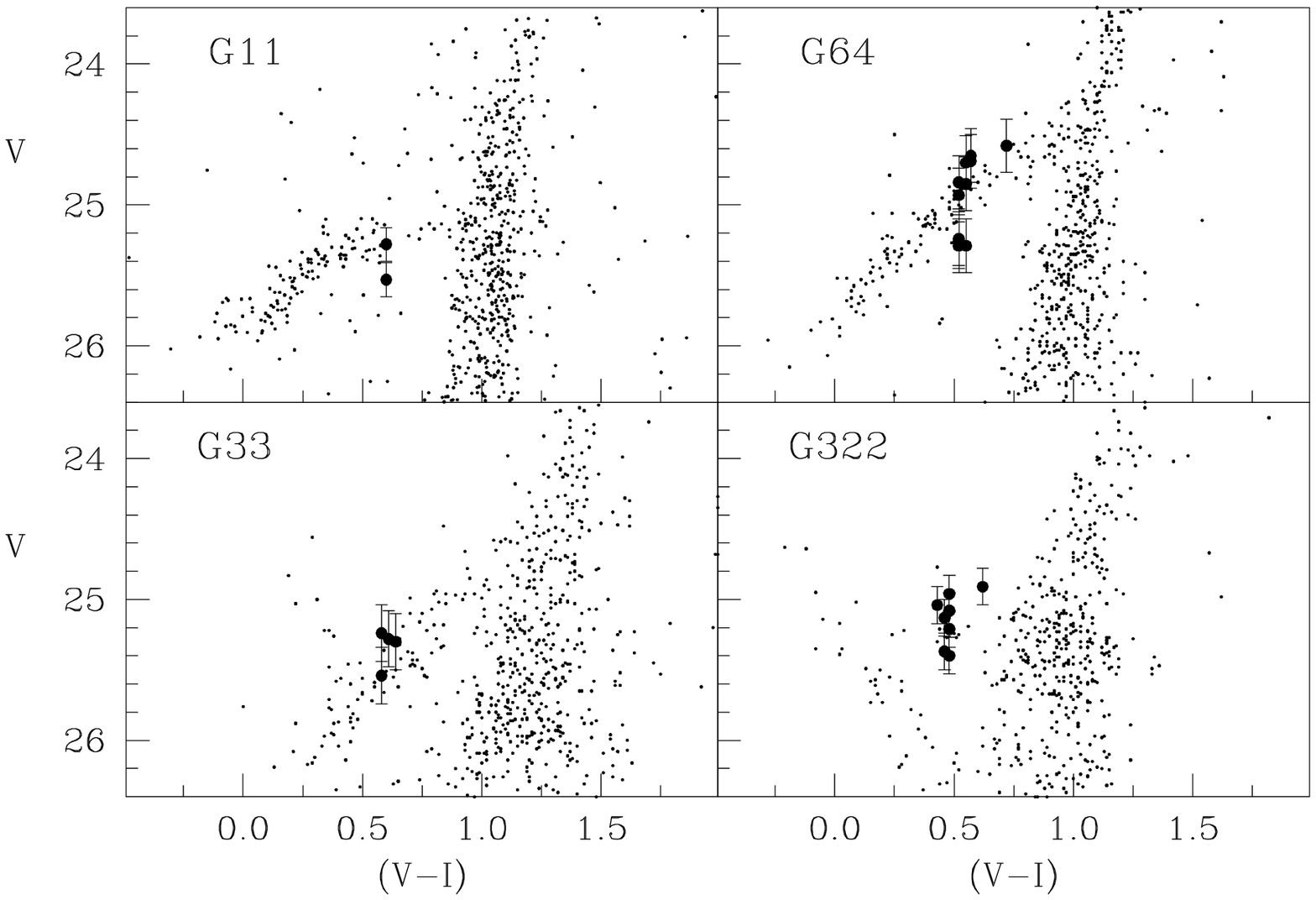}{FIG.~3.\
Location of the candidate variable stars on the CM diagram of the 
four clusters G11, G64, G33 and G322. See text for details.}

By relaxing somewhat the selection criteria (for example by accepting that 
the data points match the templates within a 2$\sigma$ error-bar) we found  
evidence for larger numbers of RR Lyrae candidates of both ab and c type
in  each of the four clusters.
In fact, an additional 14, 6, 15 and 4 candidate variables, whose 
classification is uncertain, are also present 
in G11, G33, G64 and G322, respectively.

All selected candidates are well within the tidal radii of the 
clusters ($\sim$ 16, 11, 12 and 10 arcsec in G11, G33, G64 and G322, 
respectively).

In Figure 3 we show the location of the candidate variables  
in the CMD's of their respective clusters.  The mean V magnitudes 
and V--I colors of the variable 
candidates are those of the respective templates, after adding to the 
mean V magnitudes the same ``vertical'' shifts used in item iii) before.

The major sources of error in the $<V>$ thus estimated are: a) the 
uncertainty in the reddening ($\Delta$E(B--V)= $\pm$20\% leads to 
$\Delta$V= $\pm$(0.06--0.14) mag); b) the intrinsic spread in the 
template $<V>$ (for our template stars $\Delta$V $\le$ 0.12 mag, 
however the intrinsic $<V>$ of the RR Lyraes in any globular cluster 
can span a rather large range, up to about 0.5 mag or more); and 
c) the photometric accuracy of the data points and consequent 
accuracy of the fit to the template, $\pm$0.10 mag.
This leads us to estimate error-bars on the individual $<V>$ values 
of $\pm$ 0.12, 0.20, 0.19 and 0.13 mag for the candidate variable 
stars in G11, G33, G64 and G322, respectively. 

Within these uncertainties, the agreement of the candidate RR Lyraes 
with their respective HBs is quite good, in particular it is significantly 
better than what is 
required by the original selection criteria,  further supporting 
their identification.

\section{Discussion and conclusions}
\label{s_discu}

Using HST archival data, we have found 2, 4, 11, and 8 candidate RR Lyrae
stars in the M31 globular clusters G11, G33, G64, and G322, respectively.
The evidence for the RR Lyrae nature of some of these candidates is
stronger than for others.  Nonetheless, the fact that good RR Lyrae 
candidates could be identified from an archival dataset which is not ideal 
for this sort of variability search provides clear evidence that 
suitable HST observations can be used to identify
RR Lyrae stars in M31 globular clusters. With additional HST data, it
should be possible to determine periods, light curves, and pulsational
properties for the cluster RR Lyrae.

The globular cluster G64 contains the largest number of candidate RR
Lyrae stars, combined with a predominantly blue HB.  This in itself is
not surprising since, in the Milky Way, clusters such as
M15 with predominantly blue HBs can have relatively large RR Lyrae
populations.  It is noteworthy, however, that the RR Lyrae candidates
in G64 (as well as the related HB) are unusually bright compared
to those in the other clusters, when corrected for a reddening of
E(B-V) = 0.17.  If the HB in G64 is not unusually bright, for example 
because of non-canonical phenomena such as high primordial helium abundance, 
mixing or rotation (Sweigart 1997), then
G64 is significantly closer to us (by $\sim$ 100 kpc) than the main
body of M31, or its reddening has been overestimated by about 0.10 mag.
 
Assuming $(m-M)_0$=24.43 for M31 (Freedman and Madore 1990), we note that 
the absolute magnitude of the RR Lyrae candidates turns out to be generally 
rather ``bright'', hence in better agreement with the ``long'' distance scale. 
However, this indication is very tentative and preliminary    
and needs more data and further study for confirmation.  

\acknowledgments 
We are indebted to G. Parmeggiani for help in the estimate of the GCs 
tidal radii,  and to M. Catelan for comments on the manuscript.
The data were taken within GO program 6671.  
This study was partially supported by MURST - Cofin00 under the 
project "Stellar Observables of Cosmological relevance". 
H.S. thanks the US National Science Foundation 
for support under grant AST9986943.



\begin{figure*}
\figurenum{2}
\epsscale{0.71}
\plotone{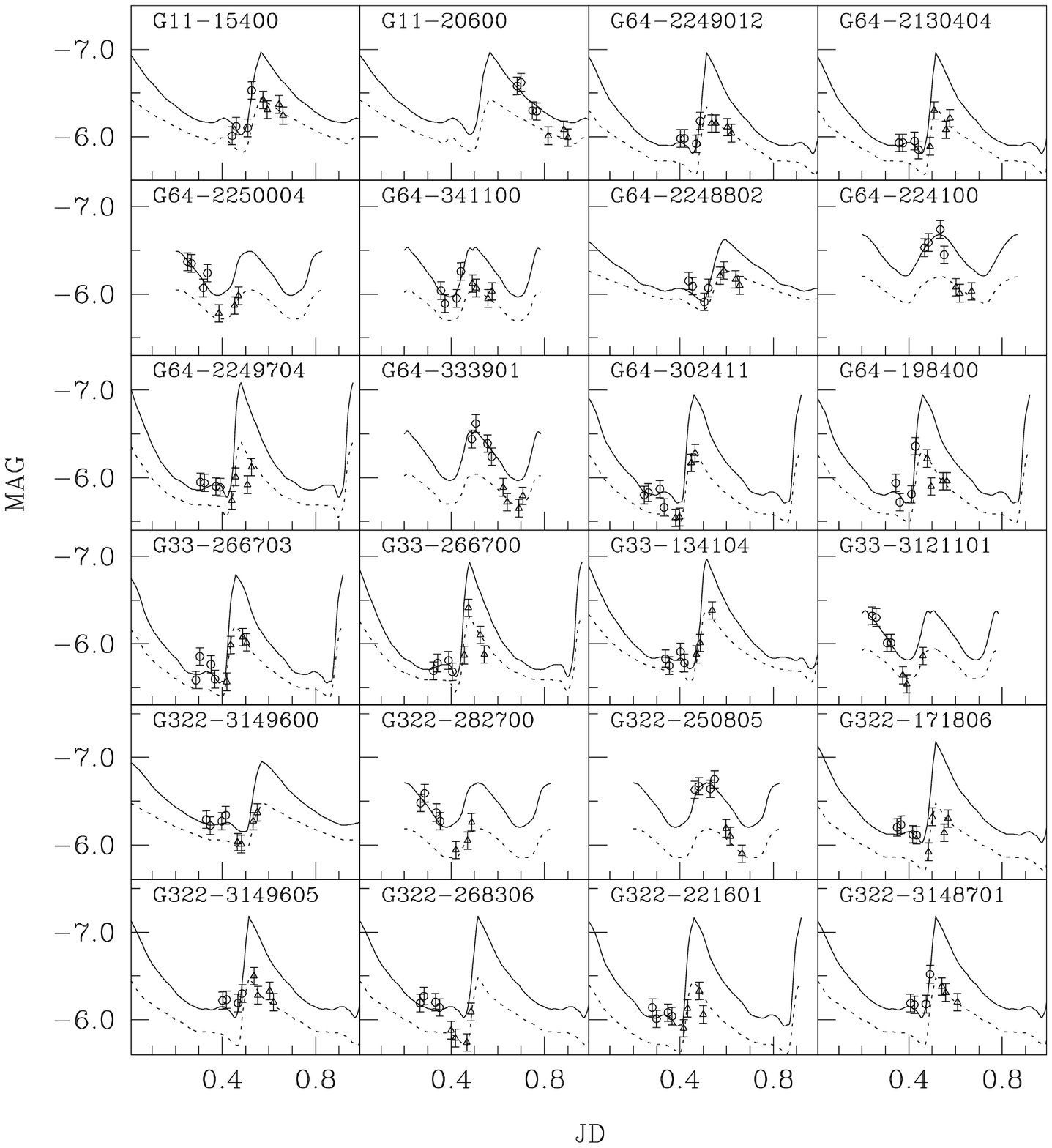}
\caption{Examples of HST V (circles) and I (triangles) data fitting 
to RR Lyrae template  V (solid lines) and I (dotted lines) light curves.
The vertical scale is magnitudes with an arbitrary zero-point, but the 
relative V and I magnitudes, both for the template light curves and for the 
observed data sets, are the real ones. The horizontal scale is Julian Days,
and the templates have been arbitrarily shifted in order to show the 
maximum light variation near the center of the frame.}
\end{figure*}

\end{document}